\documentclass[aps,prd,preprint,showkeys]{revtex4}
\usepackage{graphicx}
\usepackage{amsfonts,amsmath,amssymb,bm}

\begin{document}

\title{Thermal photon production in Au+Au collisions: viscous corrections in two different hydrodynamic formalisms}

\author{J. Peralta-Ramos}
\email{jperalta@ift.unesp.br}

\affiliation{Instituto de F\'isica Te\'orica, Universidade Estadual Paulista, 
Rua Doutor Bento Teobaldo Ferraz 271 - Bloco II, 01140-070 S\~ao Paulo, Brazil}

\author{M. S. Nakwacki}
\email{sole@iafe.uba.ar}
\affiliation{Instituto de Astronom\'ia, Geof\'isica e Ci\^encias Atmosf\'ericas, Universidade de S\~ao Paulo, 
Rua do Mat\~ao 1226, Cidade Universit\'aria, 05508-090 S\~ao Paulo, Brazil}

\date{\today}

\begin{abstract}
We calculate the spectra of produced thermal photons in Au+Au collisions taking into account the nonequilibrium contribution to photon production due to finite shear viscosity. The evolution of the fireball is modeled by second-order as well as by divergence-type 2+1 dissipative hydrodynamics, both with an ideal equation of state and with one based on Lattice QCD that includes an analytical crossover. 
The spectrum calculated in the divergence-type theory is considerably enhanced with respect to the one calculated in the second-order theory, the difference being entirely due to differences in the viscous corrections to photon production. Our results show that the differences in hydrodynamic formalisms are an important source of uncertainty in the extraction of the value of $\eta/s$ from measured photon spectra. The uncertainty in the value of $\eta/s$ associated with different hydrodynamic models used to compute thermal photon spectra is larger than the one occurring in matching hadron elliptic flow to RHIC data. 
\end{abstract}


\maketitle

\section{Introduction}

Experiments of ultrarelativistic heavy ion collisions performed at BNLs Relativistic Heavy Ion Collider (RHIC) create a hot and dense medium of quarks, antiquarks and gluons called the quark-gluon plasma (QGP) \cite{revs,luzum2,songthesis,phenixwhite}. The understanding of the transport properties of the QGP, and the nature of confinement and the phase transition from this system to interacting hadrons is a central topic of modern
high-energy physics. One of the most important discoveries at RHIC is the large elliptic flow in non-central Au+Au collisions, which is a clear indication of collective behavior. This implies that the QGP has a
very low viscosity-to-entropy ratio $\eta/s$ not much larger than the AdS/CFT lower bound $1/4\pi$ \cite{ads}. By now, it is generally agreed that the QGP thermalizes on times $\leq$ 2.5 fm/c and that it behaves as a fluid with one of the lowest viscosity-to-entropy ratios ever observed in nature $\eta$/s $\leq$ 0.5 \cite{revs,luzum2,songthesis,bound,luzum,nosprc}.

The evolution of the fireball created at RHIC has been described efficiently using relativistic hydrodynamics \cite{revs,luzum2,songthesis,phenixwhite,libro}. Ideal hydrodynamics has been partly successful in explaining the
observed collective flow at low transverse momentum and in central collisions \cite{ideal}. 
However, it is difficult to fit the data with it when a realistic equation of state (EoS) is used; see \cite{songthesis,phenixwhite}. 

The relativistic generalization of the Navier-Stokes equation is plagued with causality and stability problems \cite{libro}, so one must use the so-called second-order theories (SOTs) \cite{israel,sonhydro}. These theories rely on an expansion of the viscous tensor in velocity gradients, neglecting all orders higher than the second. They are unreliable in situations where these gradients are strong, and indeed they are known to fail, for example, in the description of strong shocks \cite{shock}. It is then valuable to develop alternative theories, not limited to weak
velocity gradients, to provide at least an estimate of the expected accuracy of the gradient
expansion. With these aims in mind, in \cite{nos} one of the authors and E. Calzetta developed an hydrodynamical description of a conformal field within the framework of the divergence-type theories
(DTT) developed by Geroch \cite{geroch} (see also \cite{calz98,liu}). DTTs do not rely on velocity
gradient expansions and therefore go beyond second-order theories. This formalism was then applied in \cite{nosprc} to study Au+Au collisions, obtaining results in good agreement with SOTs and with data on elliptic flow. It was found that the nonequilibrium correction to the particle distribution function (which is obtained from Grad's quadratic ansatz) is considerably larger in the DTT. This fact introduces significant uncertainty in the values of $\eta/s$ that can be inferred by matching the result of different dissipative hydrodynamic theories to hadron multiplicity, $<p_T>$ and elliptic flow data.

In this paper we focus on the spectra of thermal photons which are produced during the evolution of the fireball created in Au+Au collisions at $\sqrt{s_{NN}}=200$ GeV. Thermal photons are produced during the entire space-time evolution of fireball, but, since they participate only in electromagnetic interactions, they have a larger mean free path compared to the transverse size of the hot and dense matter created in nuclear collisions. Therefore, photons created in the interior of the plasma pass through the surrounding matter without any interaction, providing information on properties of bulk matter and not only on its surface. For these reasons, the emission of photons has become a useful signature of the QGP and is currently being intensively studied. See \cite{phenix,revph,processes}. 

Thermal photon spectra have been studied within the framework of ideal hydrodynamics (see \cite{revph} and references therein) and using the Israel-Stewart formalism \cite{dus2,duscorr,bhatt,bhatt1,bhatt2}. Recently, Baeuchle and Bleicher \cite{bleich} have calculated photon spectra using a hydro-kinetic hybrid approach in combination with the Ultra-relativistic Quantum Molecular Dynamics (UrQMD) transport model in 3+1 dimensions \cite{dum}. Moreover, the radial and elliptic flows, as well as the spectra and the correlations of photons are becoming a useful tool to measure or to further constrain the value of $\eta/s$ of the nuclear matter created at RHIC, as has been shown by several recent works \cite{dus2,duscorr,bhatt1,bhatt,bhatt2,v2ph,correl,gale}.

Our purpose here is to compare the thermal photon spectra obtained using the SOT and the DTT 2+1 hydrodynamics to model the evolution of the fireball, both for an ideal and a realistic EoS which includes a QGP-hadron analytic crossover as suggested by Lattice QCD calculations. To this aim, we calculate the spectra of produced thermal photons in Au+Au collisions considering the processes of Compton scattering, $q\bar{q}$-annihilation and bremsstrahlung in the QGP phase, and $\pi\pi \rightarrow \rho \gamma$, $\pi\rho \rightarrow \pi \gamma$ and $\rho \rightarrow \pi \pi \gamma$ in the hadron phase. 

The nonequilibrium contribution to photon production due to finite shear viscosity (both in the QGP and hadronic phases) is taken into account through Grad's quadratic ansatz. Very recently Bhatt, Mishra and Sreekanth \cite{bhatt2} have calculated photon spectra including viscous corrections during the QGP phase using Israel-Stewart hydrodynamics and taking into account both shear and bulk viscosity, while Dusling \cite{duscorr} has calculated the nonequilibrium correction to photon production due to Compton scattering and $q\bar{q}$ annihilation at leading-log order. We will compare our results to those of \cite{duscorr,bhatt2} later on. 

We find that the use of a realistic equation of state significantly enhances the photon spectra, in line with the results of previous calculations (see for instance \cite{revph}). The spectrum calculated in the DTT turns out to be considerably larger than the one calculated in the SOT. The difference in the spectra calculated in both formalisms is entirely due to differences in the nonequilibrium correction to photon production, which is considerably larger in the DTT. 
Our results indicate that differences in hydrodynamic formalisms constitute an important source of uncertainty in the precise determination of $\eta/s$ from observables, such as photon spectra, that depend on the nonequilibrium distribution function. 

It has been shown before \cite{nosprc,kin} that the nonequilibrium correction to the distribution function has also a strong influence on hadronic observables. The calculation of photon and hadron observables using different hydrodynamic models and its matching to data may therefore provide a way of constraining the form of the nonequilibrium correction to the distribution function and thus help improve the description of the freeze-out process \cite{kin}.  

We note that in our calculations we neglect bulk viscosity, which is known from Lattice QCD simulations to become significant near the critical temperature \cite{bulklatt}. It has been shown in \cite{bulkh} that bulk viscosity influences the space-time evolution of the fireball created in heavy-ion collisions, thus modifying the thermal photon spectra and increasing photon production \cite{bhatt}. The distribution function gets an additional correction coming from the bulk viscosity, which also modifies the spectra and puts severe limitations to the reliability of Grad's quadratic ansatz \cite{bhatt2,kin}. 
Moreover, we do not take into account prompt photons from hard scatterings of partons in the initial nuclei \cite{gordon,revph} nor jet-medium photons \cite{gale} and we only consider thermal photons. For these reasons, we do not attempt to carry out a comparison with RHIC data \cite{phenix}, but we focus instead on the comparison between different viscous hydrodynamic formalisms and on the influence on thermal photon spectra of the viscous correction to the distribution function. 

The paper is organized as follows. In section \ref{th} we briefly describe the hydrodynamic formalisms used to model the fireball evolution, and then describe the processes taken into account in the calculation of photon spectra, including viscous corrections. We present and discuss our results in section \ref{res}, and conclude in section \ref{sum}.

\section{Theoretical setup}
\label{th}

\subsection{Evolution of the fireball}
\label{hydro}

In this section we present the hydrodynamic equations of the SOT and of the DTT for boost-invariant flow in $2+1$ dimensions, for a conformal fluid. We will consider the SOT developed \cite{sonhydro}, which is based on conformal invariance and generalizes the Israel-Stewart formalism. We employ Milne coordinates defined by proper time $\tau=\sqrt{t^2-z^2}$ and rapidity $\psi=\textrm{arctanh}(z/t)$. The fluid velocity is $\vec{u}=(u^\tau,u^x,u^y,0)$ and is normalized as $u_\mu u^\mu=1$. 

The stress-energy tensor for dissipative relativistic hydrodynamics is 
\begin{equation}
\begin{split}
T^{\mu\nu} &=\rho u^\mu u^\nu - p \Delta^{\mu\nu} + \Pi^{\mu\nu} ~~~ \textrm{with} \\
\Delta^{\mu\nu} &= g^{\mu\nu}-u^\mu u^\nu
\end{split}
\end{equation}
where $\rho$ and $p$ are the energy density and the pressure in the local rest frame, and $\Pi^{\mu\nu}$ is the viscous shear tensor which is transverse ($u_\mu \Pi^{\mu\nu}=0$), traceless and symmetric. The tensor $\Delta^{\mu\nu}$ is the spatial projector orthogonal to $u^\mu$. For a conformal fluid we have $T^\mu_\mu=0$, so $\rho=3p$ and the bulk viscosity vanishes. 

In what follows, Latin indices stand for transverse coordinates $(x,y)$, $D^\mu$ is the geometric covariant derivative, $D=u_\mu D^\mu$ and $\nabla^\mu=\Delta^{\mu\nu}D_\nu$ are the comoving time and space derivatives, respectively, $\Gamma^\alpha_{\beta\gamma}$ are the Christoffel symbols and $<\ldots>$ denote the spatial, symmetric and traceless projection of a tensor. The hydrodynamic equations read
\begin{equation}
\begin{split}
(\rho+p)Du^i &= c_s^2 (g^{ij}\partial_j \rho - u^i u^\alpha \partial_\alpha \rho) - \Delta^i_\alpha D_\beta \Pi^{\alpha\beta} \\
D\rho &= -(\rho+p)\nabla_\mu u^\mu + \Pi^{\mu\nu}\sigma_{\mu\nu}
\end{split}
\label{conseq}
\end{equation}
where $c_s^2 = \partial p/\partial \rho$ is the speed of sound, $\sigma^{\mu\nu}=\nabla^{<\mu}u^{\nu>}$ is the first-order shear tensor, and 
\begin{equation}
\begin{split}
D_\beta \Pi^{\alpha\beta} &= \Pi^{i\alpha}\partial_\tau \frac{u_i}{u_\tau} + \frac{u_i}{u_\tau}\partial_\tau\Pi^{i\alpha} + \partial_i \Pi^{i\alpha} \\
&~ + \Gamma^\alpha_{\beta\gamma}\Pi^{\beta \gamma} + \Gamma^\beta_{\beta\gamma}\Pi^{\alpha\gamma} ~.
\end{split}
\label{dpiformal}
\end{equation}

In the SOT, the evolution of the shear tensor is given by \cite{sonhydro}
\begin{equation}
\begin{split}
\partial_\tau \Pi^{i\alpha} &= -\frac{4}{3u^\tau}\Pi^{i\alpha}\nabla_\mu u^\mu - \frac{1}{\tau_\pi u^\tau}\Pi^{i\alpha} + \frac{\eta}{\tau_\pi u^\tau} \sigma^{i\alpha} \\
&~ - \frac{\lambda_1}{2\tau_\pi \eta^2 u^\tau}\Pi^{<i}_\mu \Pi^{\alpha> \mu} - \frac{u^i\Pi^\alpha_\mu + u^\alpha \Pi^i_\mu}{u^\tau}Du^\mu \\
&~ -\frac{u^j}{u^\tau}\partial_j \Pi^{i\alpha} 
\end{split}
\label{dpi}
\end{equation}
where $\eta$ is the shear viscosity and $(\tau_\pi,\lambda_1)$ are second-order transport coefficients.  

In a DTT, the description of nonequilibrium hydrodynamic states requires the introduction of a new tensor $\xi^{\alpha\gamma}$ which is symmetric, traceless and vanishes in equilibrium \cite{geroch}. The evolution for $\xi^{\mu\nu}$ reads \cite{nos,nosprc}
\begin{equation}
\begin{split}
\partial_\tau \xi^{i\alpha} &= -\frac{2}{3u^\tau}\xi^{i\alpha}\nabla_\mu u^\mu - \frac{1}{\tau_\pi u^\tau}\xi^{i\alpha} + \frac{1}{\tau_\pi u^\tau} \sigma^{i\alpha} \\
&~ - \frac{\lambda_1}{3\tau_\pi \eta u^\tau}\xi^{<i}_\mu \xi^{\alpha> \mu} - \frac{u^i\xi^\alpha_\mu + u^\alpha \xi^i_\mu}{u^\tau}Du^\mu \\
&~ -\frac{u^j}{u^\tau}\partial_j \xi^{i\alpha} ~.
\end{split}
\label{dxi}
\end{equation}
and the shear tensor is calculated from the nonequilibrium tensor $\xi^{\alpha\gamma}$ as follows 
\begin{equation}
\Pi^{\mu\nu} = \eta \xi^{\mu\nu} -\frac{\lambda_1 \tau_{\pi} T^4}{3\eta} (\xi^{\mu\alpha}\xi^\nu_\alpha - \frac{1}{3}\Delta^{\mu\nu}\xi^{\alpha\gamma}\xi_{\alpha\gamma}) ~.
\end{equation}

As independent variables we choose $(\rho,u^x,u^y,\Pi^{xx},\Pi^{xy},\Pi^{yy})$ for the SOT and $(\rho,u^x,u^y,\xi^{xx},\xi^{xy},\xi^{yy})$ for the DTT. The $\tau$ component of the velocity follows from normalization, while the other nontrivial components of $\Pi^{\mu\nu}$ (and of $\xi^{\mu\nu}$) follow from the transversality and tracelessness conditions. We use vanishing initial transverse velocity and shear tensor, and set the initialization time $\tau_0=$ 1 fm/c and the initial and freeze-out temperatures $T_i=333$ MeV and $T_f=140$ MeV, respectively. 
The initial energy density profile is calculated using Glauber's model \cite{glau} (see \cite{nosprc} for details).
In all calculations we set the impact parameter $b=$ 3 fm, and we use a 13 fm $\times$ 13 fm transverse plane and values for the second-order transport coefficients corresponding to a strongly-coupled $\cal{N}=$ 4 Super-Yang Mills (SYM) plasma \cite{ads,sonhydro}: $\tau_\pi = 2(2-\ln 2)\eta/(sT)$ and $\lambda_1=\eta/(2\pi T)$, where $s$ is the entropy density. In order to calculate the photon spectra, we take the critical temperature to be $T_c=$ 170 MeV which is consistent with recent Lattice QCD calculations (see \cite{tclatt} and references therein).

\begin{figure}[htb]
\scalebox{1}{\includegraphics{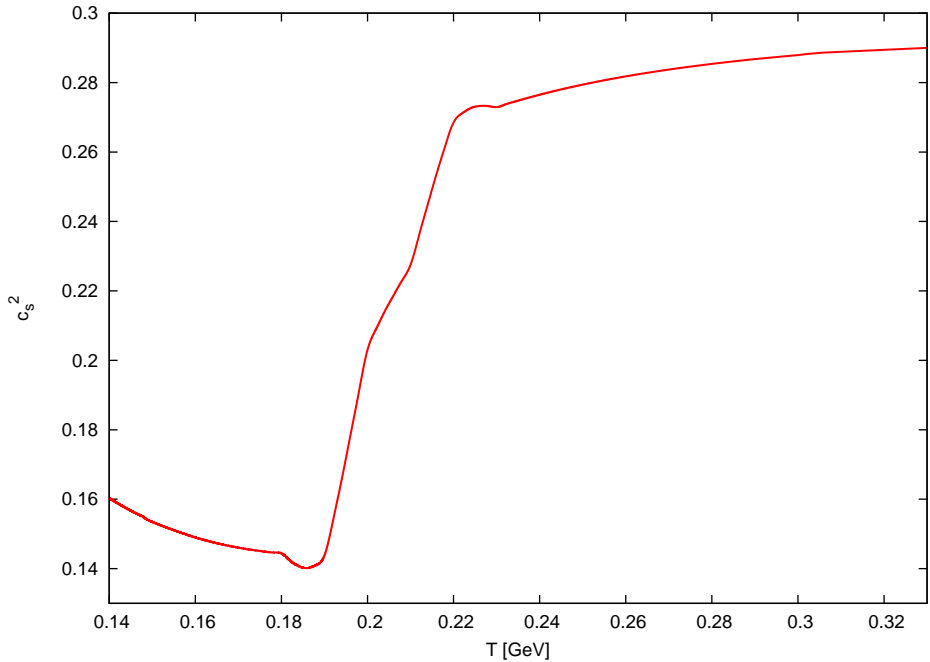}}
\vspace{1cm}
\caption{(Color online) Square of the velocity of sound $c_s^2$ as a function of temperature corresponding to the EoS of Laine and Schr\"{o}der \cite{laine}. This EoS connects a high-order weak-coupling perturbative QCD calculation at high temperatures to a hadron resonance gas at low temperatures via an analytic crossover.}
\label{fig1}
\end{figure}

The set of hydrodynamic equations must be closed with an equation of state. We employ two different EoS: the one computed by Laine and Schr\"{o}der \cite{laine}, which connects a high-order weak-coupling perturbative QCD calculation at high temperatures to a hadron resonance gas at low temperatures via an analytic crossover (as suggested by Lattice QCD calculations \cite{aoki}), and an ideal one with $p=\rho/3$. In Fig. \ref{fig1} we reproduce the realistic EoS (which we will call LQCD EoS in what follows), showing the square of the velocity of sound $c_s^2=\partial p/\partial \rho$ as a function of temperature. 
We note that using the LQCD EoS and these values for $\tau_0$, $\tau_\pi$ and $\lambda_1$, data on Kaon elliptic flow, $<p_T>$ and total multiplicity can be consistently reproduced both within the SOT and the DTT \cite{luzum,luzum2,nosprc}.  

\subsection{Photon production}
\label{ph}
In this paper we consider thermal photon production due to the processes of Compton scattering, $q\bar{q}$ annihilation and bremsstrahlung in the QGP phase, and $\pi\pi \rightarrow \rho \gamma$, $\pi\rho \rightarrow \pi \gamma$ and $\rho \rightarrow \pi \pi \gamma$ in the hadron phase. For a review on photon production in the context of heavy-ion collisions see \cite{revph}.

The Compton scattering and $q\bar{q}$ annihilation contribution is \cite{com}
\begin{equation}
E\frac{dN}{d^4xd^3p}=\frac{1}{2\pi^2}\alpha\alpha_s(\sum_fe^2_f)T^2e^{-E/T}\ln\bigg(\frac{cE}{\alpha_sT}\bigg)
\end{equation}
where $c\sim$ 0.23, $\alpha=1/137$ and \cite{as}
\begin{equation}
 \alpha_s(T) =\frac{6\pi}{(33-2N_f)\ln(8T/T_c)}
\end{equation}
where $N_f$ is the number of quark flavors. In the summation $f$ stands for quark flavor while $e_f$ is the electric charge of the quark in units of the charge of the electron.
The rate due to $q\bar{q}$ annihilation with an additional scattering in the thermalized medium is given by 
\begin{equation}
E\frac{dN}{d^4xd^3p}=\frac{8}{3\pi^5}\alpha\alpha_s(\sum_fe^2_f)ETe^{-E/T}(J_T-J_L) ~,
\end{equation}
where $J_T\sim$ 1.11, $J_L \sim$ 1.06.

The rate of photon production due to bremsstrahlung reads \cite{brem}
\begin{equation}
\begin{split}
E\frac{dN}{d^4xd^3p}&=\frac{8}{\pi^5}\alpha\alpha_s(\sum_fe^2_f)\frac{T^4}{E^2}e^{-E/T}(J_T-J_L)
\bigg[3\zeta(3)+\frac{\pi^2E}{6T} \\
&+\frac{E^2}{T^2}\ln 2 +4\textrm{Li}_3(-e^{-E/T})+2\frac{E}{T}\textrm{Li}_2(-e^{-E/T})-\frac{E^2}{T^2}
\ln(1+e^{-E/T})\bigg]
\end{split}
\end{equation}
where $\zeta$ is the zeta function and Li$_m=\sum_{n=1}^\infty z^n/n^m$ are polylogarithmic functions. 

In order to calculate the photon spectrum from the fireball when the LQCD EoS is used, one has to know also the photon production rate from the hot hadron gas since photons will also be emitted from this thermal phase following the QGP. We use the simple estimate for the photon production rate given by Steffen and Thoma \cite{hhg}, which reproduces the sum of production rates for the processes $\pi\pi \rightarrow \rho \gamma$, $\pi\rho \rightarrow \pi \gamma$ and $\rho \rightarrow \pi \pi \gamma$ calculated in \cite{hhgprev}. The estimate for the production rate reads
\begin{equation}
E\frac{dN}{d^4xd^3p} \simeq ~4.8 T^{2.15} e^{-1/(1.35 ET)^{0.77}} e^{-E/T} ~.
\end{equation}

The photon production rates just described are calculated with the equilibrium distribution function within Boltzmann's approximation, hence the factor $e^{-E/T}$. However, it is well known that viscous effects lead to the modification of the thermal distribution functions which in turn modify observables such as hadronic elliptic flow \cite{kin,luzum,nosprc,revs}. These modifications may also have significant observational effects on the photon spectra as has been shown recently \cite{duscorr,bhatt2}. It is therefore interesting to incorporate the viscous correction to the distribution function and asses its impact on final photon spectra. To this end, we calculate the photon production rates using the nonequilibrium distribution function coming from Grad's ansatz which is given by \cite{revs}
\begin{equation}
f(x^\mu,p^\mu) = f_0+f_0(1-f_0)\frac{p^\mu p^\nu\Pi_{\mu\nu}}{2T^2 (\rho + p)} ~~,
\label{dis}
\end{equation}
with $f_0$ the Fermi-Dirac function, and compare with the results obtained by using the equilibrium distribution function. It is known that Grad's ansatz becomes unrealible at high transverse momentum $p_T \gtrsim $ 3$-$4 GeV \cite{nosprc,df,bhatt2} and for this reason we will restrict to lower values of $p_T$ in what follows.  
 
The total thermal photon spectrum is obtained by integrating the sum of the production rates for each process given above  over the space-time evolution of the fireball created in the collision. It is given by 
\begin{equation}
 \bigg(\frac{dN}{d^2p_TdY}\bigg)=\int d\vec{x}\int^{\tau_2}_{\tau_1}\int^{Y}_{-Y}E\frac{dN}{d^4xd^3p}
\end{equation}
where $\tau_{1,2}$ are the initial and final time one is interested in, $Y$ is the rapidity of the nuclei, $d\vec{x}=(dx,dy)$ and $p_T$ is the transverse momentum. Note that the photon energy in the comoving frame is given by $p_T \textrm{cosh}(Y-Y')$. 

\section{Results}
\label{res}

We now present our results on photon spectra, in all cases at midrapidity $Y=0$.

\begin{figure}[htb]
\scalebox{1}{\includegraphics{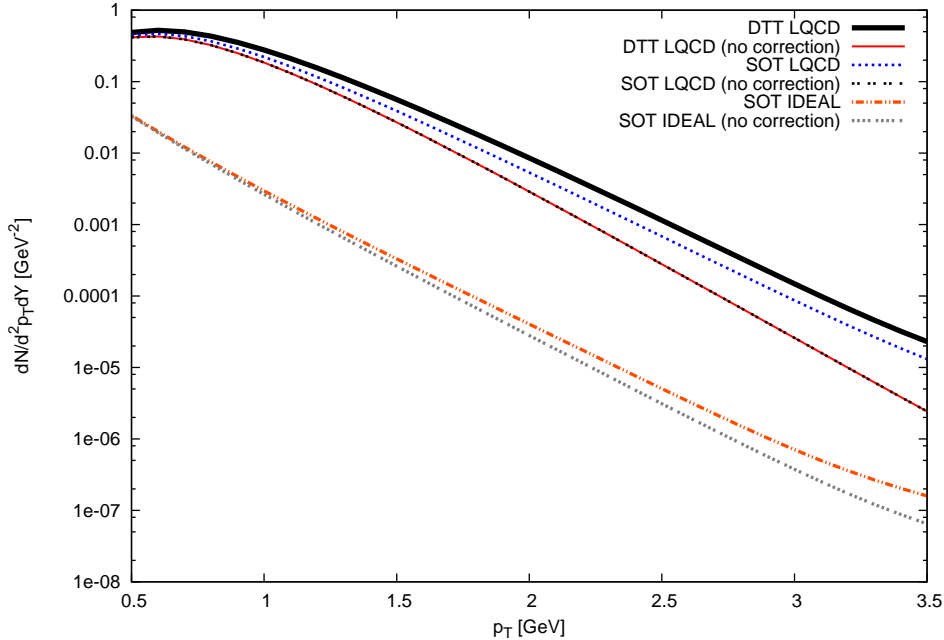}}
\vspace{1cm}
\caption{(Color online) Thermal photon spectra calculated with the SOT and the DTT for $\eta/s =$ 0.3, both for the ideal and the LQCD equations of state.}
\label{figspec}
\end{figure}
In Fig. \ref{figspec} we show the total thermal photon spectra calculated with the SOT and the DTT for $\eta/s =$ 0.3, both for the ideal EoS and for the LQCD EoS, and including or not the nonequilibrium correction to the distribution function. We do not show the spectra calculated with the DTT for the ideal EoS because it is very similar to that obtained in the SOT. 

It is seen that the spectra strongly depend on the EoS used, being two orders of magnitude smaller in the case of the ideal EoS. This is a consequence of the fact that the hydrodynamic evolution of the plasma is slower when the realistic EoS is used and hence more thermal photons are produced (see \cite{revph,bhatt,bhatt2}). Besides, as already shown in \cite{bhatt2,duscorr}, the effect of the nonequilibrium correction to the distribution function is to harden the spectra. 

Going over to the comparison between the SOT and DTT, it is seen that the differences in the spectra arise entirely from the nonequilibrium corrections that become important only at $p_T >$ 1 GeV. The spectrum obtained from the DTT is significantly larger, and the difference in spectra increases with increasing $p_T$. As we will show in what follows, this is due to the fact that the nonequilibrium corrections to the distribution function are larger in the DTT, and therefore the viscous correction to photon production is larger too (for a discussion of this point in the context of hadronic elliptic flow see \cite{nosprc}). 

\begin{figure}[htb]
\scalebox{1}{\includegraphics{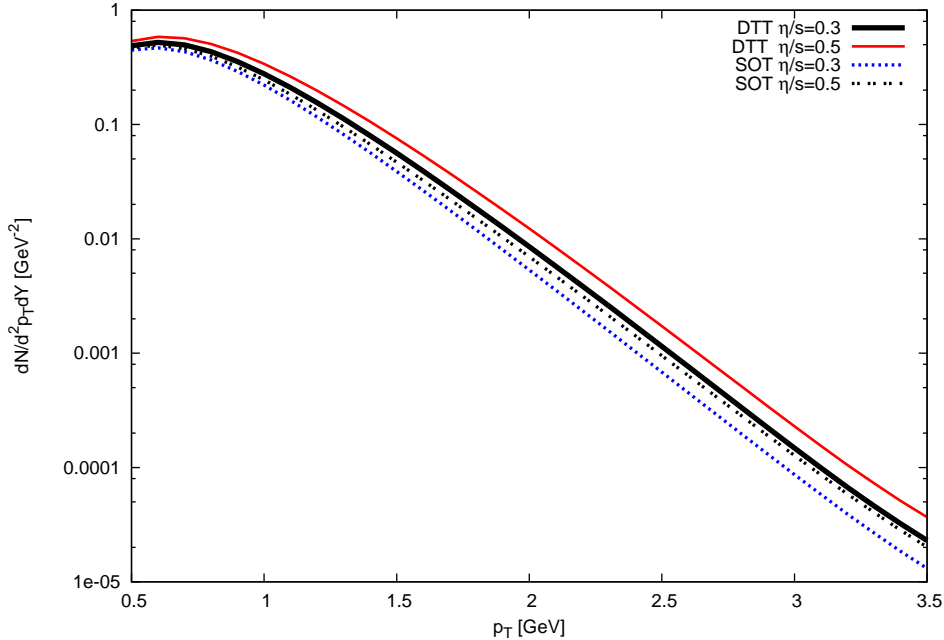}}
\vspace{1cm}
\caption{(Color online) Thermal photon spectra calculated with the SOT and the DTT for $\eta/s =$ 0.3 and $\eta/s =$ 0.5.}
\label{speceta}
\end{figure}

In Fig. \ref{speceta} we compare the total photon spectra obtained with the LQCD EoS in the DTT and the SOT with $\eta/s=$ 0.3 and $\eta/s=$ 0.5. It is seen that the spectra is strongly dependent on the value of $\eta/s$, in agreement with the results of \cite{duscorr}. Moreover, the difference in spectra corresponding to $\eta/s=$ 0.3 and $\eta/s=$ 0.5 is slightly larger in the DTT, showing that the latter formalism is more sensitive to changes in the value of the viscosity-to-entropy ratio. It is interesting to note that the spectra obtained in the SOT with $\eta/s=$ 0.5 is very similar to that obtained in the DTT with $\eta/s$ = 0.3, which clearly indicates that differences in hydrodynamic formalisms constitute a significant source of uncertainty in the extraction of $\eta/s$ from photon spectra data. A similar conclusion was found in \cite{nosprc} with respect to the charged hadron elliptic flow, although in that case differences between the SOT and the DTT have a smaller influence on the value of $\eta/s$ that can be extracted from data by matching to hydrodynamic results.


\begin{figure}[htb]
\scalebox{1}{\includegraphics{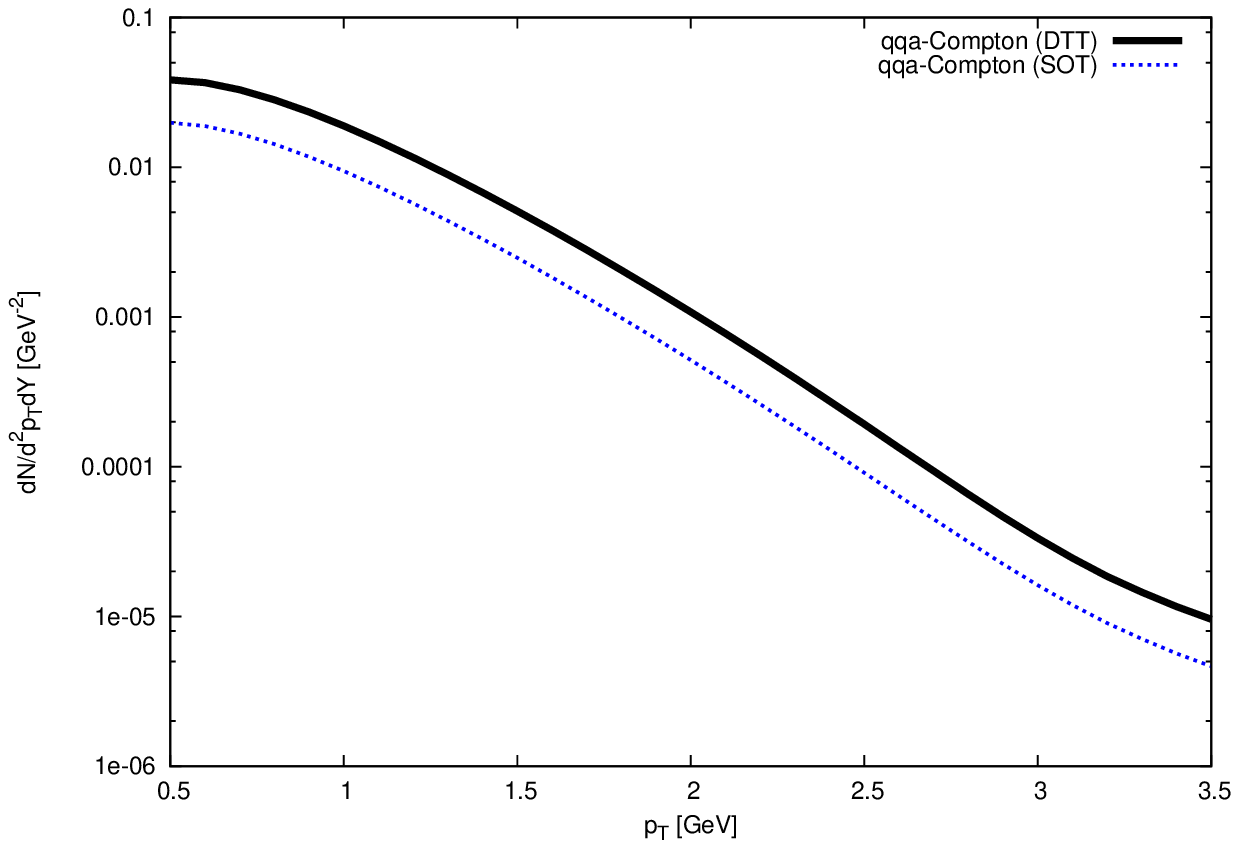}}
\vspace{1cm}
\caption{(Color online) Nonequilibrium contribution to photon spectra calculated in the SOT and in the DTT taking into account only $q\bar{q}$-annihilation and Compton scattering, for $\eta/s =$ 0.3.}
\label{figqq}
\end{figure}
\begin{figure}[htb]
\scalebox{1}{\includegraphics{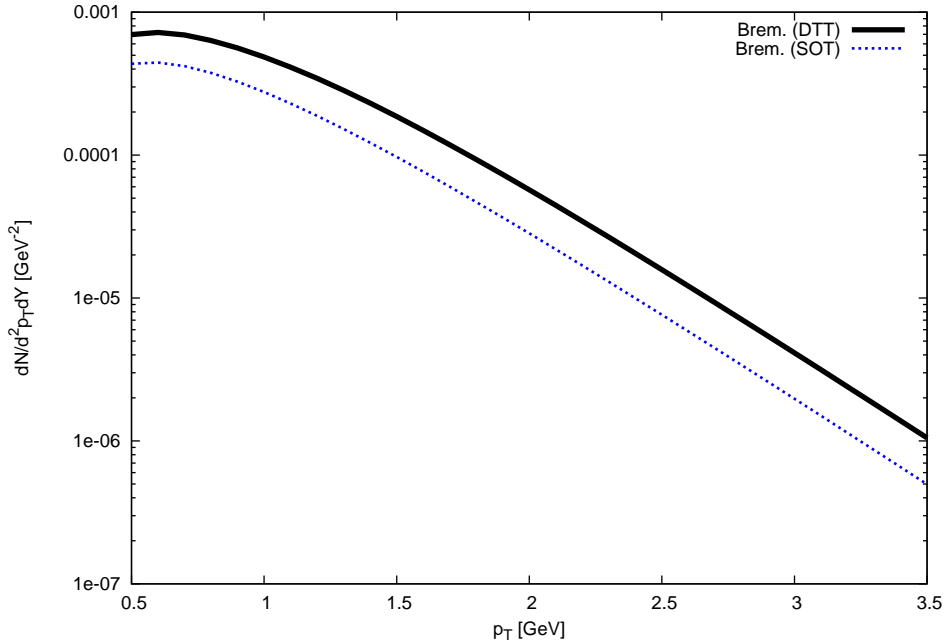}}
\vspace{1cm}
\caption{(Color online) Nonequilibrium contribution to photon spectra calculated in the SOT and in the DTT taking into account only bremsstrahlung, for $\eta/s =$ 0.3.}
\label{figB}
\end{figure}
\begin{figure}[htb]
\scalebox{1}{\includegraphics{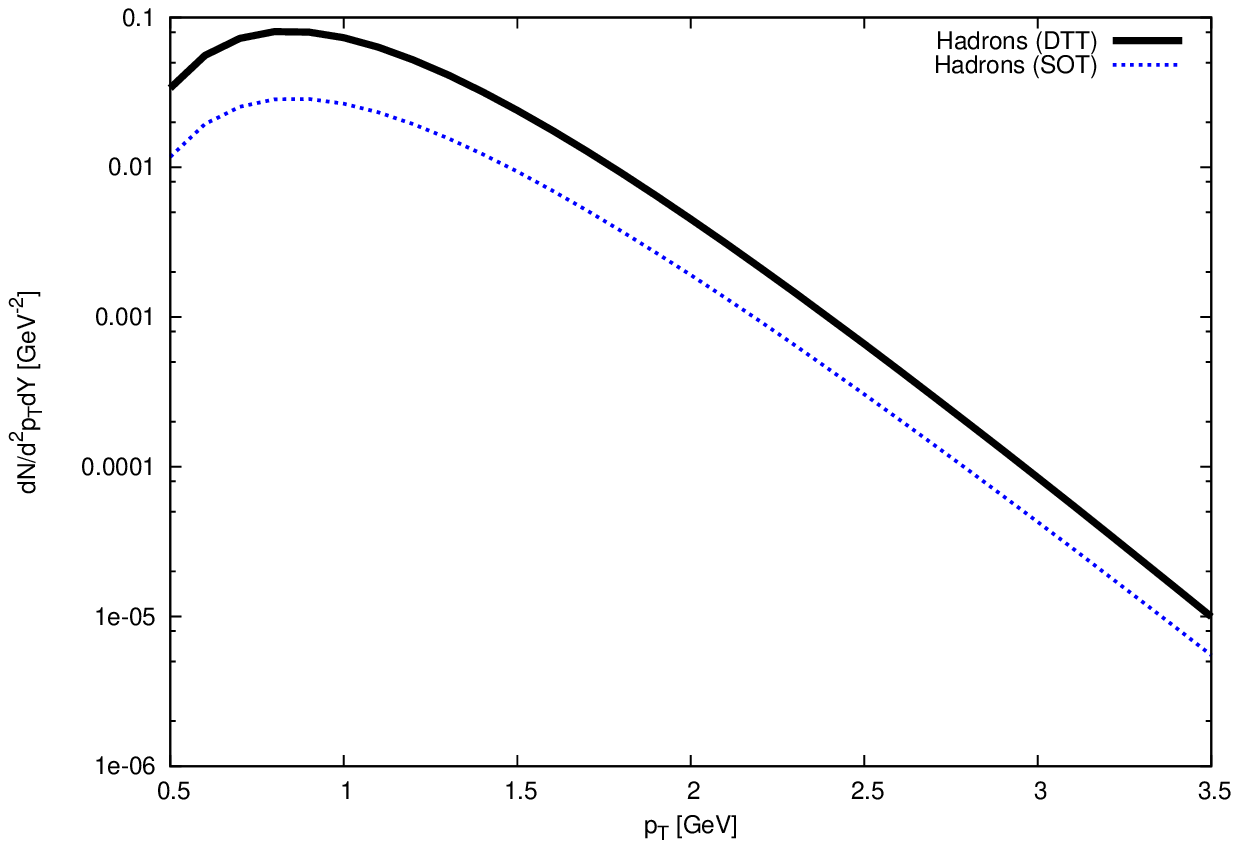}}
\vspace{1cm}
\caption{(Color online) Nonequilibrium contribution to photon spectra calculated in the SOT and in the DTT taking into account only photon production in the hadronic phase, for $\eta/s =$ 0.3.}
\label{figH}
\end{figure}
In order to determine the influence of the nonequilibrium contribution on the different processes considered, in Figs. \ref{figqq}-\ref{figH} we show this contribution integrated over the evolution of the fireball in the SOT and the DTT for $\eta/s =$ 0.3, taking into account only the processes of $q\bar{q}$-annihilation and Compton scattering, bremsstrahlung, or considering only the hadronic phase, respectively. It is seen that in all three cases the nonequilibrium correction is significantly larger in the DTT. The nonequilibrium correction to the spectra is larger in the hadronic phase at low $p_T$, as is the difference between the DTT and the SOT as it can be seen from Fig. \ref{figH}. At larger values of $p_T$ the nonequilibrium correction is larger in the QGP phase. The difference between the spectra calculated in the DTT and in the SOT is almost independent of $p_T$ for the $q\bar{q}$-annihilation/Compton scattering and bremsstrahlung processes, while in the hadronic phase it decreases with increasing transverse momentum.

\begin{figure}[htb]
\scalebox{1}{\includegraphics{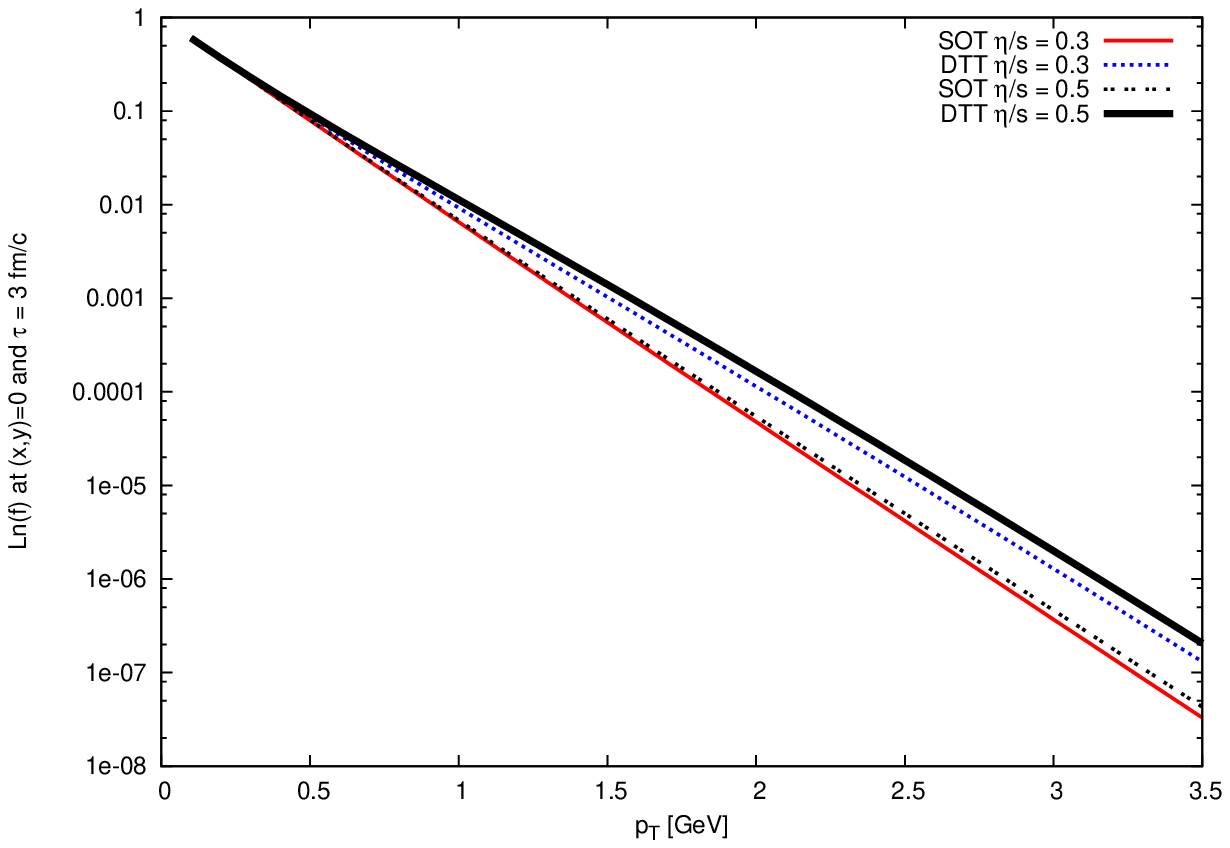}}
\vspace{1cm}
\caption{(Color online) Nonequilibrium distribution function evaluated at the center of the fireball and at $\tau=$ 3 fm/c as a function of $p_T$ calculated in the DTT and the SOT with $\eta/s$ = 0.3 and $\eta/s$ = 0.5.}
\label{ftau}
\end{figure}
The differences between the nonequilibrium contribution to photon spectra in the SOT and in the DTT can be understood from Fig. \ref{ftau}, where we show the nonequilibrium distribution function of Eq. (\ref{dis}) evaluated at $(x,y)=0$ and at $\tau=$ 3 fm/c as a function of $p_T$ for the DTT and the SOT with $\eta/s$ = 0.3 and $\eta/s$ = 0.5. Several interesting features, some of which were already discussed when showing our results on spectra, can be seen from the figure. The difference between the DTT and the SOT becomes appreciable at $p_T >$ 0.5 GeV. Beyond this value it is seen that the distribution function in the DTT is considerably larger than the one in the SOT. Moreover, the effect of increasing the value of $\eta/s$ is larger in the DTT, which reflects the fact that the viscous correction to $f(x^\mu,p^\mu)$ is larger in the DTT.

From these results we conclude that the nonequilibrium contribution to photon production strongly modifies the spectra, even at low transverse momenta, resulting in harder spectra (this is in line with the results of \cite{duscorr,bhatt2}).  Moreover, we have found that the difference between hydrodynamic formalisms leads to significant uncertainty in the extraction of $\eta/s$ from thermal photon spectra.
  
It has been shown before \cite{nosprc,kin} that the nonequilibrium correction to the distribution function has also a strong influence on hadronic observables, especially on the elliptic flow. Therefore, the calculation of photon and hadronic observables using different hydrodynamic models may provide a way of determining or at least constrain, albeit phenomenologically, the form of the nonequilibrium correction to the distribution function. This may help improve the theoretical description of the freeze-out process, which is currently based on Grad's quadratic ansatz (see \cite{kin} for recent developments).

\section{Summary}
\label{sum}

We have calculated the spectra of thermal photons produced in Au+Au collisions taking into account the processes of Compton scattering, $q\bar{q}$-annihilation and bremsstrahlung in the QGP phase, and $\pi\pi \rightarrow \rho \gamma$, $\pi\rho \rightarrow \pi \gamma$ and $\rho \rightarrow \pi \pi \gamma$ in the hadron phase. The calculation was done using two dissipative hydrodynamics formalisms to model the 2+1 evolution of the fireball, namely the second-order theory based on conformal invariance and a divergence-type theory. We have focused on determining the influence of the nonequilibrium correction to the distribution function on photon production in both formalisms. We have also compared the results obtained with an ideal and a realistic equation of state based on Lattice QCD that includes an analytical crossover between the quark-gluon plasma and a resonance hadron gas.

In agreement with previous studies (\cite{bhatt,bhatt1,bhatt2,revph}) we have found that the use of a realistic equation of state enhances the photon spectra, implying that the QGP-hadron crossover can not be neglected in the investigation of thermal photons. Besides, the spectra calculated in the DTT is significantly enhanced with respect to the one calculated in the SOT. The difference is entirely due to differences in the nonequilibrium corrections to photon production in both formalisms. 

Our results point to the conclusion that differences in dissipative hydrodynamic formalisms are a significant source of uncertainty in the precise determination of $\eta/s$ from data, particularly from those observables, such as the one studied here, that depend on the nonequilibrium distribution function. Comparing our results with those obtained for the  charged hadron elliptic flow and its matching to RHIC data \cite{nosprc}, we find that the difference in hydrodynamic formalisms used to calculate thermal photon spectra leads to a larger uncertainty in the possible extraction of $\eta/s$ from measured photon spectra. 
  
It is interesting to note that the dependence of hadron or photon observables on the nonequilibrium correction to the distribution function, which leads to uncertainty in the extraction of $\eta/s$ from data, may in turn serve to constrain the form of the nonequilibrium distribution function. In this way one could gain some insight on the relation between kinetic theory and hydrodynamic models as well as on possible generalizations of Grad's quadratic ansatz, and help improve the theoretical description of the freeze-out process in heavy-ion collisions \cite{kin}. For this to be feasible, the theoretical uncertainties present in the hydrodynamic description of heavy-ion collisions, for example those regarding the initial conditions, must be under control.

In this paper we have limited ourselves to the case of vanishing bulk viscosity. As mentioned in the Introduction, Lattice QCD results indicate that the bulk viscosity becomes peaked at the critical temperature. It would be interesting to include bulk viscosity in the DTT to determine its influence on thermal photon spectra in this hydrodynamic model. 

\begin{acknowledgments}
We thank Gast\~ao Krein, Esteban Calzetta and V. Sreekanth for valuable comments and suggestions. This work has been supported by FAPESP (Brazil).
\end{acknowledgments}


\begin{thebibliography}{99}
\bibitem{revs} P. Romatschke, Int. J. Mod. Phys. E {\bfseries 19}, 1 (2010); U. Heinz, arXiv:0901.4355 [nucl-th]; A. Muronga, Phys. Rev. C {\bfseries 69}, 034903 (2004). 
\bibitem{luzum2} M. Luzum, {\it Relativistic Heavy Ion Collisions: Viscous Hydrodynamic Simulations and Final State Interactions}, PhD Thesis (University of Washington, 2009), 0908.4100 [nucl-th].
\bibitem{songthesis} H. Song, {\it Causal Viscous Hydrodynamics for Relativistic Heavy Ion Collisions}, PhD Thesis (Ohio State University, 2009), 0908.3656 [nucl-th].
\bibitem{phenixwhite} K. Adcox et al (PHENIX Collaboration), Nuclear Physics A {\bfseries 757}, 184 (2005).
\bibitem{ads} P. K. Kovtun, D. T. Son, and A. O. Starinets, Phys. Rev. Lett. {\bfseries 94}, 111601 (2005); D. T. Son and A. O. Starinets, Annual Review of Nuclear and Particle Science, {\bfseries 57}, 95 (2007).
\bibitem{bound} S. Gavin and M. Abdel-Aziz, Phys. Rev. Lett. {\bfseries 97}, 162302 (2006); H. J. Drescher, A. Dumitru, C. Gombeaud, and J. Y. Ollitrault, Phys. Rev. C {\bfseries 76}, 024905 (2007); A. K. Chaudhuri, Phys. Lett. B {\bfseries 681}, 418 (2009).
\bibitem{luzum} M. Luzum and P. Romatschke, Phys. Rev. C {\bfseries 78}, 034915 (2008); Erratum-ibid.C {\bfseries 79}, 039903 (2009). 
\bibitem{nosprc} J. Peralta-Ramos and E. Calzetta, Phys. Rev. C {\bfseries 82}, 054905 (2010).
\bibitem{libro} E. Calzetta and B.-L. Hu, {\it Nonequilibrium Quantum Field Theory} (Cambridge University Press, Great Britain, 2008).
\bibitem{ideal} D. Teaney, J. Lauret, and E. V. Shuryak, Phys. Rev. Lett. {\bfseries 86}, 4783 (2001); P. Huovinen, P. F. Kolb, U. W. Heinz, P. V. Ruuskanen, and
S. A. Voloshin, Phys. Lett. B {\bfseries 503}, 58 (2001); T. Hirano and K. Tsuda, Phys. Rev. C {\bfseries 66}, 054905 (2002); P. F. Kolb and R. Rapp, Phys. Rev. C {\bfseries 67}, 044903 (2003); Piotr Bozek and Iwona Wyskiel, Phys. Rev. C {\bfseries 79}, 044916 (2009). 
\bibitem{israel} W. Israel, Ann. Phys. (NY) {\bfseries 100}, 310 (1976); W. Israel, and J. Stewart, Ann. Phys. (N.Y.) {\bfseries 118}, 341 (1979).
\bibitem{sonhydro} R. Baier, P. Romatschke, D. T. Son, A. O. Starinets, and M. A. Stephanov, J. High Energy Phys. {\bfseries 04}, 100 (2008); S. Bhattacharyya, V. E. Hubeny, S. Minwalla, and M. Rangamani, J. High Energy Phys. {\bfseries 02}, 45 (2008); M. Natsuume and T. Okamura, Phys. Rev. D {\bfseries 77}, 066014 (2008); Erratum-ibid. D {\bfseries 78}, 089902 (2008).
\bibitem{shock} I. Bouras, E. Molnar, H. Niemi, Z. Xu, A. El, O. Fochler, C. Greiner and D.H. Rischke, arXiv:1006.0387 [hep-ph]; ibid, Nucl. Phys. A {\bfseries 830}, 741 (2009); S. Khlebnikov, M. Kruczenski, and G. Michalogiorgakis, arXiv:1004.3803.
\bibitem{nos} J. Peralta-Ramos and E. Calzetta, Phys. Rev. D {\bfseries 80}, 126002 (2009). 
\bibitem{geroch} R. Geroch and L. Lindblom, Phys. Rev. D {\bfseries 41}, 1855 (1990).
\bibitem{liu} I.-S. Liu, I. Muller, and T. Ruggeri, Ann. Phys. (N.Y.) {\bfseries 169}, 191 (1986); T. Ruggeri, in {\it Relativistic Fluid Dynamics}, Lecture Notes in Mathematics Vol. 1385, Eds. A. Anile and Y. Choquet-Bruhat (Springer-Verlag, Germany, 1989).
\bibitem{calz98} E. Calzetta, Class. Quant. Grav. {\bfseries 15}, 653 (1998).
\bibitem{revph} T. Peitzmann, and M. H. Thoma, Phys. Rep. {\bfseries 364}, 175 (2002).
\bibitem{processes} D. de Florian, and G. F. R. Sborlini, arXiv:1011.0486v1 [hep-ph]; M. A. Betemps, and M. V. T. Machado, arXiv:1010.4738; W.-L. Sang, and Y.-Q. Chen, Phys. Rev. D {\bfseries 81}, 034028 (2010); T. P. Stavreva, and J. F. Owens, Phys. Rev. D {\bfseries 79}, 054017 (2009); L. Bhattacharya, and P. Roy, Phys. Rev. C {\bfseries 78}, 064904 (2008). 
\bibitem{phenix} A. Adare et al. (PHENIX Collaboration), Phys. Rev. Lett. {\bfseries 104}, 132301 (2010).
\bibitem{duscorr} K. Dusling, Nucl. Phys. A {\bfseries 839}, 70 (2010).
\bibitem{dus2} K. Dusling and I. Zahed, arXiv:0911.2426 [nucl-th].
\bibitem{bhatt} J. R. Bhatt, H. Mishra, and V. Sreekanth, arXiv:1005.2756 [hep-ph].
\bibitem{bhatt1} J. R. Bhatt, and V. Sreekanth, Int. J. Mod. Phys. E {\bfseries 19}, 299 (2010).
\bibitem{bhatt2} J. R. Bhatt, H. Mishra, and V. Sreekanth, arXiv:1011.1969 [hep-ph].
\bibitem{bleich} B. Baeuchle and M. Bleicher, arXiv:1008.2332 [nucl-th]; ibid, arXiv:1008.2338 [nucl-th]; ibid, Phys. Rev. C {\bfseries 81}, 044904 (2010). 
\bibitem{dum} A. Dumitru, M. Bleicher, S. A. Bass, C. Spieles, L. Neise, H. Stoecker, and W. Greiner, Phys. Rev. C {\bfseries 57} 3271 (1998).
\bibitem{v2ph} F.-M. Liu, T. Hirano, K. Werner, and Y. Zhu, arXiv:0902.1303 [hep-ph]; R. Chatterjee, E. Frodermann, U. Heinz, and D. K. Srivastava, Phys. Rev. Lett. {\bfseries 96}, 202302 (2006); R. Chatterjee, and D. K. Srivastava, Phys. Rev. C {\bfseries 79}, 021901 (2009); P. Mohanty, J. K. Nayak, J.-E. Alam, and S. K. Das, Phys. Rev. C {\bfseries 82}, 034901 (2010).
\bibitem{gale} C. Gale, S. Turbide, E. Frodermann, and U. Heinz, J. Phys. G: Nucl. Part. Phys. {\bfseries 35}, 104119 (2008).
\bibitem{correl} S. De, D. K. Srivastava, and R. Chatterjee, arXiv:1008.1475 [nucl-th]; G.-Y. Qin, J. Ruppert, C. Gale, S. Jeon, and G. D. Moore, Phys. Rev. C {\bfseries 80}, 054909 (2009).
\bibitem{kin} M. Luzum and J.-Y. Ollitrault, arXiv:1004.2023 [nucl-th]; G. S. Denicol, T. Kodama, T. Koide, and Ph. Mota, Phys. Rev. C {\bfseries 80}, 064901 (2009); A. Monnai and T. Hirano, Phys. Rev. C {\bfseries 80}, 054906 (2009); K. Dusling, G. Moore, and D. Teaney, Phys. Rev. C {\bfseries 81}, 034907 (2010); E. Calzetta, and J. Peralta-Ramos, Phys. Rev. D {\bfseries 82}, 106003 (2010).
\bibitem{bulklatt} H. B. Meyer, Phys. Rev. Lett. {\bfseries 100}, 162001 (2008).
\bibitem{bulkh} H. Song, and U. W. Heinz, Phys. Rev. C {\bfseries 81}, 024905 (2010); J. W. Li, Y. G. Ma, G. L. Ma, Chin. Phys. B {\bfseries 18}, 4786 (2009); K. Rajagopal, and N. Tripuraneni, JHEP {\bfseries 1003}, 018 (2010); G. S. Denicol, T. Kodama, and T. Koide, arXiv:1002.2394 [nucl-th]; P. Bozek, arXiv:0911.2397 [nucl-th]; A. Monnai and T. Hirano, Nucl. Phys. A {\bfseries 830}, 471 (2009).
\bibitem{gordon} L. E. Gordon, and W. Vogelsang, Phys. Rev. D {\bfseries 48}, 3136 (1993).
\bibitem{glau} P. F. Kolb, U. W. Heinz, P. Huovinen, K. J. Eskola, and
K. Tuominen, Nucl. Phys. A {\bfseries 696}, 197 (2001).
\bibitem{tclatt} S. Borsanyi, Z. Fodor, C. Hoelbling, S. D. Katz, S. Krieg, C. Ratti, and K. K. Szabo, arXiv:1005.3508 [hep-lat].
\bibitem{laine} M. Laine and Y. Schr\"{o}der, Phys. Rev. D {\bfseries 73}, 085009 (2006).
\bibitem{aoki} Y. Aoki, G. Endrodi, Z. Fodor, S. D. Katz and K. K. Szabo, Nature {\bfseries 443}, 675 (2006); P. Huovinen and P. Petreczky, Nucl. Phys. A {\bfseries 837}, 26 (2010); P. Huovinen, Nucl. Phys. A {\bfseries 761}, 296 (2005); T. Hirano and M. Gyulassy, Nucl. Phys. A {\bfseries 769}, 71 (2006); U. W. Heinz, J. Phys. G {\bfseries 31}, S717 (2005); P. Huovinen, Eur. Phys. J. A {\bfseries 37}, 121 (2008).
\bibitem{com} J. Kapusta, P. Lichard, and D. Seibert, Phys. Rev. D {\bfseries 44}, 2774 (1991); C. T. Traxler, H. Vija, and M. H. Thoma, Phys. Lett. B {\bfseries 346}, 329 (1995).
\bibitem{as} C. T. Traxler, and M. H. Thoma, Phys. Rev. C {\bfseries 53}, 1348 (1996).
\bibitem{brem} P. Aurenche, F. Gelis, H. Zaraket, and R. Kobes, Phys. Rev. D {\bfseries 58}, 085003 (1998).
\bibitem{hhg} F.D. Steffen, and M.H. Thoma, Phys. Lett. B {\bfseries 510}, 98 (2001).
\bibitem{hhgprev} H. Nadeau, J.I. Kapusta, and P. Lichard, Phys. Rev. C {\bfseries 45} 3034 (1992), and Erratum-ibid C.
{\bfseries 47} 2426 (1993); C. Song, Phys. Rev. C {\bfseries 47} 2861 (1993). 
\bibitem{df} H. Song and U. Heinz, Phys. Rev. C {\bfseries 77}, 064901 (2008); A. K. Chaudhuri, Phys. Rev. C {\bfseries 74}, 044904 (2006); M. Luzum and J.-Y. Ollitrault, arXiv:1004.2023 [nucl-th]; K. Dusling and D. Teaney, Phys. Rev. C. {\bfseries 77}, 034905 (2008).


\end{thebibliography}
\end{document}